\documentclass{sig-alternate}
\usepackage{times}
\usepackage{paralist}
\newcommand{\superscript}[1]{\ensuremath{^{\textrm{#1}}}}
\def\sharedaffiliation{\end{tabular}\newline\begin{tabular}{c}}

\newlength{\wideitemsep}
\let\olditem\item
\renewcommand{\item}{\setlength{\itemsep}{\wideitemsep}\olditem}

\usepackage{fancybox}
\usepackage{graphicx}
\graphicspath{{figures/}}
\DeclareGraphicsExtensions{.pdf,.jpeg,.png}
\usepackage{amsmath}
\usepackage{subfigure}
\usepackage{caption}
\usepackage{fancyvrb}
\usepackage{color}
\usepackage{algorithmic}
\usepackage{array}
\usepackage{hyperref}
\usepackage{mdwmath}
\usepackage{mdwtab}
\usepackage{booktabs}
\usepackage{listings}
\usepackage{eqparbox}
\usepackage{stfloats}
\usepackage{url}
\usepackage{comment}
\usepackage{framed}
\usepackage{ifthen}
\usepackage{balance}
\usepackage[turnon]{notes}
\usepackage{multirow}
\usepackage{xspace}
\usepackage{listings}

\renewcommand{\baselinestretch}{0.99}

\newcommand{\rom}[1]{\uppercase\expandafter{\romannumeral #1\relax}}

\newcommand{\etal}{\hbox{\emph{et al.}}\xspace}
\newcommand{\eg}{\hbox{\emph{e.g.,}}\xspace}
\newcommand{\ie}{\hbox{\emph{i.e.}}\xspace}

\newcommand{\wrt}{\hbox{\emph{w.r.t.}}\xspace}

\newcommand{\etc}{\hbox{\emph{etc.}}\xspace}

\newcommand{\DP}{DP\xspace}

\newcommand{\nbf}{\hbox{${\cal N}\hspace{-0.01in}{\cal B}\hspace{-0.01in}{\cal F}$}\xspace}
\newcommand{\sta}{\hbox{${\cal S}\hspace{-0.01in}{\cal B}\hspace{-0.01in}{\cal F}$}\xspace}
\newcommand{\dpm}{\hbox{${\cal D}\hspace{-0.01in}{\cal P}$}\xspace}
\newcommand{\ngm}{\$gram\xspace}

\newcommand{\ngmwt}{\$gram+wType\xspace}
\newcommand{\fbg}{\hbox{\sc FindBugs}\xspace}
\newcommand{\pmd}{\hbox{\sc pmd}\xspace}

\newcommand{\gh}{\hbox{Github}\xspace}
\newcommand{\atm}{\hbox{Atmosphere}\xspace}
\newcommand{\el}{\hbox{Elasticsearch}\xspace}
\newcommand{\fdk}{\hbox{FDK}\xspace}
\newcommand{\netty}{\hbox{Netty}\xspace}
\newcommand{\prs}{\hbox{Presto}\xspace}
\newcommand{\drb}{\hbox{Derby}\xspace}
\newcommand{\luc}{\hbox{Lucene}\xspace}
\newcommand{\oj}{\hbox{OpenJPA}\xspace}
\newcommand{\qpd}{\hbox{Qpid}\xspace}
\newcommand{\wck}{\hbox{Wicket}\xspace}

\newcommand{\tableshrink}[1]{\vspace{-0.25in}}

\definecolor{gray50}{gray}{.5}
\definecolor{gray40}{gray}{.6}
\definecolor{gray30}{gray}{.7}
\definecolor{gray20}{gray}{.8}
\definecolor{gray10}{gray}{.9}
\definecolor{gray05}{gray}{.95}

\newlength\Linewidth
\def\findlength{\setlength\Linewidth\linewidth
\addtolength\Linewidth{-4\fboxrule}
\addtolength\Linewidth{-3\fboxsep}
}
\newenvironment{examplebox}{\par\begingroup
   \setlength{\fboxsep}{5pt}\findlength
   \setbox0=\vbox\bgroup\noindent
   \hsize=0.95\linewidth
   \begin{minipage}{0.95\linewidth}\normalsize}
    {\end{minipage}\egroup
    \textcolor{gray20}{\fboxsep2.5pt\fbox
     {\fboxsep5pt\colorbox{gray05}{\normalcolor\box0}}}
    \endgroup\par\noindent
    \normalcolor\ignorespacesafterend}

\newcounter{RQCounter}
\newcounter{RQACounter}

\newcommand{\RQ}[2]{%
\refstepcounter{RQCounter} \label{#1}
 \begin{center}	
  \begin{examplebox}
   \textbf{RQ\arabic{RQCounter}.}~#2
  \end{examplebox}	 
 \end{center}
}

\newcommand{\RQA}[2]{%
\refstepcounter{RQACounter} \label{#1}
\vspace{0.1in} \noindent\textbf{RQ\arabic{RQACounter}.~#2 \vspace{0.05in}}

}

\newcommand{\RS}[2]{%
\begin{framed}%
\filbreak
\textbf{Result {\ref{#1}}:~}{\emph {#2}}%
\end{framed}
}

\definecolor{javared}{rgb}{0.6,0,0} 
\definecolor{javagreen}{rgb}{0.25,0.5,0.35} 
\definecolor{javapurple}{rgb}{0.5,0,0.35} 
\definecolor{javadocblue}{rgb}{0.25,0.35,0.75} 

\lstdefinestyle{customc}{
  belowcaptionskip=\baselineskip,
  breaklines=true,
  xleftmargin=\parindent,
  language=java,
  showstringspaces=false,
  basicstyle=\scriptsize\ttfamily,
  keywordstyle=\bfseries\color{javapurple},
  commentstyle=\itshape\blue,
  belowskip=-10pt,
  aboveskip=-5pt
}

\newcommand\Red[1]{\textcolor[rgb]{1.00,0.00,0.00}{\textbf{#1}}}
\newcommand\red[1]{\textcolor[rgb]{1.00,0.00,0.00}{#1}}

\newcommand\blue[1]{\textcolor[rgb]{0.00,0.00,1.00}{{#1}}}

\newcommand\dkgreen[1]{\textcolor[rgb]{0.0,0.6,0}{\textbf{#1}}}

\def\wu{$^{\dag}$}
\def\wg{\superscript{\ddag}}
\def\wn{$^{\natural}$}

\newcommand{\nl}{\ensuremath{\cal{N\!L}}\xspace}
\newcommand{\ngram}{$n$gram\xspace}
\newcommand{\ngrams}{$n$grams\xspace}

\newcommand{\cgram}{\emph{\$gram}\xspace}
\begin{document}
\title{On the ``Naturalness'' of Buggy Code}\vspace{-2ex}
\numberofauthors{1}
\author{
  \alignauthor Baishakhi Ray\wu\titlenote{Baishakhi Ray and Vincent Hellendoorn are both first authors, and contributed equally to the work.}\hfill Vincent Hellendoorn\wg\raisebox{8pt}{$\ast$} \hfill Zhaopeng Tu\wn \hfill Connie Nguyen\wu\\ 
  \hfill Saheel Godhane\wu\hfill Alberto Bacchelli\wg\hfill Premkumar Devanbu\wu\hfill
  \sharedaffiliation
  \begin{tabular}{ccc}
    \affaddr{University of California, Davis} 			& \affaddr{Huawei Technologies Co. Ltd.}& \affaddr{Delft University of Technology} \\
    \affaddr{Davis, CA 95616, USA}            			& \affaddr{Sha Tin, Hong Kong, China}& \affaddr{Delft, Netherlands} \\
    {\small \email{\small{\{bairay,ptdevanbu,srgodhane,cmnguyen\}}@ucdavis.edu}}	&  \email{\small{tuzhaopeng}@gmail.com}& \email{\small{\{V.J.Hellendoorn,A.Bacchelli\}}@tudelft.nl}
  \end{tabular}
}


\maketitle

\vspace{-2.5em}

\begin{abstract}

Real software, the kind working programmers produce by the kLOC to solve
real-world problems, tends to  be ``natural'', like speech or natural language; it  
tends to be highly repetitive and predictable.
Researchers have captured this \emph{naturalness of software} through statistical models
and used them to good effect in suggestion engines, porting tools, coding standards checkers, and idiom miners.  
This suggests that code that appears improbable, or surprising, to a good
statistical language model is ``unnatural'' in some sense, and thus possibly
suspicious.
In this paper, we investigate this hypothesis.
We consider a large corpus of \emph{bug fix commits} (ca.~8,296),  from 10 different Java projects, 
and we focus on its language statistics, evaluating the naturalness of buggy code and the corresponding
fixes.
We find that code with bugs tends to be more entropic (\ie unnatural), becoming less so as bugs are fixed.
Focusing on highly entropic lines is similar in cost-effectiveness 
to some well-known static bug finders (PMD, FindBugs) and ordering warnings from these bug finders using
an entropy measure improves the cost-effectiveness of inspecting code implicated in warnings. This suggests that entropy may be a valid language-independent and simple way to complement the 
effectiveness of PMD or FindBugs, and that search-based bug-fixing
methods may benefit from using entropy both for fault-localization and searching for
fixes. 

\end{abstract}





\section{Introduction}\label{sec:intro}

Communication is ordinary, everyday human behavior,  something we do \emph{naturally}. This ``natural'' linguistic behavior is characterized by efficiency and fluency, rather than creativity. Most natural language (NL) is both repetitive
and predictable, thus enabling humans to communicate reliably \& efficiently
in potentially noisy and dangerous situations. This repetitive property, \ie \emph{naturalness}, of spoken and written NL has been exploited in the field of NLP: Statistical language models (from hereon: \emph{language models}) have been employed to capture it,
 and then use to good effect in speech recognition, translation, spelling correction, \etc 

As it turns out, so it is with code! People also write code using repetitive, predictable, forms: Recent work~\cite{Hindle:2012:ICSE} showed that code is amenable to the same kinds of language modeling as NL, and language models have been used to good effect in code suggestion~\cite{Hindle:2012:ICSE, Raychev:2014:PLDI, Tu:2014:FSE, Franks:2015:ICSE}, 
cross-language porting~\cite{Nguyen:2013:FSE, Nguyen:2014:ICSEa, Nguyen:2014:ICSEb, Karaivanov:2014:SPLASH}, coding standards~\cite{Allamanis:2014:FSE}, idiom mining~\cite{Allamanis:2014:FSE2}, and code de-obfuscation~\cite{Raychev:2015:POPL}. 
Since language models are useful in these tasks,  they are capturing some property of how code is supposed to be. This raises an interesting question: \emph{What does it mean when a code fragment is considered improbable by these models?} 

Language models assign higher naturalness to code (tokens, syntactic forms, \etc) frequently encountered during training, and lower naturalness to code  rarely or never seen. In fact, prior work~\cite{Campbell:2014:MSR} showed that syntactically incorrect code is flagged as improbable by language models. However, by restricting ourselves to code that occurs in repositories, we still encounter unnatural, yet syntactically correct code; why?
We hypothesize that \emph{unnatural code is more likely to be wrong}, thus, language models actually help zero-in on potentially defective code; in this paper, we explore this. 

To this end, we consider a large corpus of 8,296 bug fix commits from 10 different projects, and we focus on its language statistics, evaluating the naturalness of defective code and whether fixes increase naturalness. Language models can rate probabilities of linguistic events at any granularity, even at the level of characters. We focus here on line-level defect analysis, giving far finer granularity of prediction than traditional defect prediction methods, which most often operate at the granularity of files or modules. 
In fact, this approach is more commensurate with static analysis or static bug-finding tools, which also 
indicate potential bugs at line-level.  For this reason, we also investigate our language model approach in contrast and in conjunction with two well-known static bug finders (namely, PMD~\cite{copeland2005pmd} and FindBugs~\cite{findbugs}).

Overall, our results corroborate our initial hypothesis that code with bugs tends to be more unnatural. In particular, the main findings of this paper are:
\vspace{-0.5em}

%


\begin{enumerate}
\item Buggy code is rated as significantly more ``unnatural'' (improbable) by language models. 
\item This unnaturalness drops significantly when buggy code is replaced by fix code. 
\item Using cost-sensitive measures, inspecting ``unnatural'' code indicated by language models works quite well: Performance is comparable to that of static bug finders FindBugs and PMD. 
\item Ordering warnings produced by the FindBugs and PMD tools, using the ``unnaturalness'' of associated code, significantly improves the performance of these tools. 
\end{enumerate}

\vspace{-0.5em}

Our experiments are mostly done with Java projects, but we have strong empirical evidence indicating that the first two findings above generalize to C as well; we hope to confirm the rest in future work.


\section{Background}
Our main goal is evaluating the degree to which defective code appears ``unnatural'' to language models, and to what extent language
models can actually enable programmers to zero-in on bugs during inspections. Furthermore, if language models can actually help direct programmers
towards buggy lines, we are interested to know how they compare against static bug-finding tools. In this section, we present 
relevant technical background and the main research questions. We begin with a brief technical background on language modeling.

\subsection{Language Modeling}
\label{sec:cache}

\noindent
\textbf{\textit{Basics.}} 
Language models are statistical models that assign a probability to every sequence of {\em words}. 
Given a code sequence $S = t_1 t_2 \dots t_N$, a language model estimates the probability of this sequence occurring as a product of a series of conditional probabilities for each token:
\begin{equation}
P(S) = P(t_1) \cdot \prod_{i=2}^{N} P(t_i | t_1, \dots, t_{i-1})
\end{equation}
Each probability $P(t_i | t_1, \dots, t_{i-1})$ denotes the chance that the token $t_i$ follows the previous tokens, the {\em prefix}, $h = t_1, \dots, t_{i-1}$.
In practice, however, the probabilities are impossible to estimate, as there is an astronomically large number of possible prefixes.
The most widely used approach to combat this problem is to use the \ngram language model, which makes a {\em Markov assumption} that the conditional probability of a token is dependent only on the $n-1$ most recent tokens.
The \ngram model places all prefixes that have the same $n-1$ tokens in the same equivalence class:
\begin{equation}
P_{\textrm{\ngram}}(t_i|h) = P(t_i | t_{i-n+1}, \dots, t_{i-1})
\end{equation}
The latter is estimated from the training corpus as the fraction of times that the prefix $t_{i-n+1}, \dots, t_{i-1}$ was followed by the token $t_i$. Note that, given a complete sentence, we can also compute each token given its epilog (the subsequent tokens), essentially computing the probability of the sentence in reverse. We make use of this approach to better identify buggy lines, as described in \ref{subsec:building}.

The \ngram language models have been shown to successfully capture the highly repetitive regularities in source code, and were applied to code suggestion tasks~\cite{Hindle:2012:ICSE}. However, the \ngram models fail to deal with a special property of software: source code is \underline{\emph{also very localized}}.
Due to module specialization and focus, code tends to take special repetitive forms in local contexts.
The  \ngram approach, rooted as it is in \nl, focuses on capturing the global regularities over the whole corpus, and neglects local regularities, thus ignoring the {\em localness of software}. To overcome this, Tu \etal~\cite{Tu:2014:FSE} introduced a cache language model to capture the localness of code.

\vspace{5pt}
\noindent
\textbf{\textit{Cache language models.}} These models (for short: \cgram) extend the traditional language models by deploying an additional {\em cache} to capture the regularities in the locality. It combines the global (\ngram) model with the local ({\em cache}) model as
\begin{equation}
P(t_i|h, cache) = \lambda \cdot P_{\textrm{\ngram}}(t_i|h) + (1- \lambda ) \cdot P_{\textrm{\em cache}}(t_i|h)
\end{equation}
$cache$ is the list of \ngrams extracted from the local context, and $P_{\textrm{\em cache}}(t_i|h)$ is estimated from the frequency with which $t_i$ followed the prefix $h$ in the {\em cache}.
To avoid hand-tuned parameters, Tu \etal~\cite{Tu:2014:FSE} replaced the interpolation weight $\lambda$ with ${\gamma}/{(\gamma+H)}$, where $H$ counts the times the prefix $h$ has been observed in the {\em cache}, and $\gamma$ is a concentration parameter between 0 and infinity.
\begin{equation}
P(t_i|h, cache) =  \frac{\gamma}{\gamma+H} \cdot P_{\textrm{\ngram}}(t_i|h) +  \frac{H}{\gamma+H} \cdot P_{\textrm{\em cache}}(t_i|h)
\label{equation-final-score}
\end{equation}
If the prefix occurs few times in the cache ($H$ is small), then the \ngram model probability will be preferred; vice versa. This setting makes the interpolation weight self-adaptive for different \ngrams.

The \ngram and cache components capture different regularities: the \ngram component captures the corpus linguistic structure,
and offers a good estimate of the mean probability of a specific linguistic event in the corpus; 
around this mean, the local probability fluctuates,  as code patterns change in different localities.  
The cache component models these local changes, and provides variance around the corpus mean for different local contexts.

We use a \cgram here to judge the ``improbability'' (measured as cross-entropy) of lines of code; the core research questions being, can cross-entropy provide a useful indication of the likely bugginess of a line of code, and how does this approach performs against/with comparable approaches, such as static bug finders. 

\subsection{Static Bug-finders (\sta)}
The goal of \sta is to use syntactic and semantic properties of source code to indicate locations of common errors, such as undefined variables and buffer overflows.
They rely on methods ranging from informal heuristic pattern-matching to formal algorithms with proven properties. These tools typically report warnings at build time; programmers can choose to fix them. Pattern-matching tools (\eg PMD and FindBugs~\cite{copeland2005pmd,findbugs}) are unsound, but fast and widely used; more formal approaches are sound, but slower. In practice all tools have false positives and/or false negatives, thus inspecting and fixing all the warnings is not always cost-effective. 

Suffice for our purposes to note here that both \sta and \cgram both (fairly imperfectly) indicate likely locations of defects; so our goal here is to compare these  rather different approaches, and see if they synergize. 
It should be noted that \cgram is quite easy to implement, since it requires only lexical information; however, as we see below it can be improved with some syntactic information. 

\begin{table*}[ht!pb]
\scriptsize
  \centering
    \begin{tabular}{lp{3cm}rrrrrr}

     Ecosystem & Project & \multicolumn{1}{c}{Study Period} & \multicolumn{1}{c}{Snapshots} & \multicolumn{1}{c}{\#Files} & \multicolumn{1}{c}{NCSL}  & \multicolumn{1}{c}{\# of Changes} & \multicolumn{1}{c}{\# of Bugs} \\
    \midrule
    \multicolumn{1}{c}{\multirow{5}[0]{*}{\gh}} & Atmosphere & May-10 to Jan-14 & 17    & 17,206 & 6,329,400 & 2,481  & 1,130 \\
    \multicolumn{1}{c}{} & Elasticsearch & Feb-10 to Jan-14 & 17    & 103,727 & 22,156,904 & 4,922  & 1,077 \\
    \multicolumn{1}{c}{} & Facebook-android-sdk (fdk) & May-10 to Dec-13 & 16    & 3,981 & 1,431,787 & 320   & 143 \\
    \multicolumn{1}{c}{} & Netty & Aug-08 to Jan-14 & 24    & 57,922 & 12,969,858 & 3,906  & 1,485 \\
    \multicolumn{1}{c}{} & Presto & Aug-12 to Jan-14 & 7     & 23,086 & 6,496,149 & 1,635  & 330 \\
    \midrule
    \multicolumn{1}{c}{\multirow{5}[0]{*}{Apache}} & Derby & Sep-04 to Jul-14 & 41    & 143,906 & 61,192,709 & 5,275  & 1,453 \\
    \multicolumn{1}{c}{} & Lucene & Sep-01  to Mar-10 & 36    & 47,270 & 11,744,856 & 2,563  & 469 \\
    \multicolumn{1}{c}{} & OpenJPA & May-06 to Jun-14 & 34    & 131,441 & 27,709,778 & 2,956  & 558 \\
    \multicolumn{1}{c}{} & Qpid  & Sep-06 to Jun-14 & 33    & 94,790 & 24,031,170 & 3,362  & 657 \\
    \multicolumn{1}{c}{} & Wicket & Sep-04 to Jun-14 & 41    & 159,332 & 28,544,601 & 10,583 & 994 \\
    \midrule 
  Overall &  & Sep-01 to Jul-14 & 266   & 782,661 &  202,607,212 & 38,003 & 8,296 \\

    \end{tabular}%
\caption{{\small{\bf Summary data of projects that are analyzed for finding all the defects including development time bugs}}}
\label{tab:study}
\end{table*}%

\subsection{Evaluating Defect Predictions}
\label{subsec:evaluating}
In our setting, we view \sta and \cgram  as two commensurate approaches to selecting lines of code to which drawing the programmers' attention as locations worthy of inspection, since they just might contain real bugs. To emphasize this similarity, from here on we refer to language model based bug prediction as  \nbf (``{\bf N}aturalness {\bf B}ug {\bf F}inder").
With either \sta or \nbf, programmers will spend effort on reviewing the code and hopefully find some defects. 
Comparing the two approaches requires a performance measure. We 
adopt a cost-based measure that has become standard~\cite{Arisholm2010Systematic}: AUCEC (Area Under the Cost-Effectiveness Curve).   AUCEC (like ROC) is a non-parametric measure, which does not depend on the defects' distribution. AUCEC assumes that the cost is the inspection effort and the payoff is the count of  bugs found. 

We normalize both to 100\%, measure the 
payoff increase as we inspect more and more lines and draw a `lift-chart' or Lorenz curve. AUCEC is the area under this curve. 
Suppose we inspect x\% code at random;  in expectation, we would find x\% of the bugs, thus yielding a diagonal line on the lift chart; so the expected AUCEC if inspecting 5\% lines at random would be 0.00125.\footnote{Calculated as 0.5 * 0.05 * 0.05. This could be normalized differently; but we consistently use this measurement, so our comparisons work.} 
Typically, inspecting 100\% code is very expensive; one could reasonably assume that 5\% or even just 1\% of the code, in a large system, could realistically be inspected; therefore, we compare AUCECs for \nbf and \sta for this much smaller proportion. 

Additionally, we investigate defect prediction performance under several \emph{credit criteria}. A prediction model is awarded credit, ranging from 0 to 1, for each line marked as defective. 
Previous work by Rahman \etal has compared \sta and \dpm (a file level statistical defect predictor) models using two types of credit: full (or optimistic) and partial (or scaled) credit \cite{rahman2014comparing}, which we adapt to line level defect prediction. The former metric awards a model one credit point for each bug iff at least one line of the bug was marked buggy by the model. Thus, it assumes that a programmer will spot a bug as soon as one of its lines is identified as such. Partial credit is more conservative: For each bug, the credit awarded to the model is  the fraction of the bug's defective lines that the model marked. Hence, partial credit assumes that the probability of a developer finding a bug is proportional to the fraction of the bug that is marked by the defect prediction model. 

\subsection{Research Questions}

At the core of our research is the question whether ``\emph{un}nat\-ural\-ness" (measured as entropy, or improbability) is indicative of poor code quality. 
The abundant history of changes (including bug fixes) in OSS projects  allows the use of standard methods~\cite{sliwerski2005changes} to find code that was
implicated in bug fixes (``buggy code'').

\RQ{rq1}{Are buggy lines less ``natural'' than non-buggy lines?}

In project histories, we can find numerous samples of bug fixes, where buggy code is replaced by bug-fix code to correct defects. Do language models rate bug-fix code as  more natural than the buggy code they replaced? This would essentially mean that the bug fix code is assigned a higher probability than the buggy code. Such a finding would also have implications for automatic, search-based bug repair: If fixes tend to have higher probability, then a good language model might provide an effective organizing principle for the search, or perhaps (if the model is generative) even generate possible candidate repairs. 

\RQ{rq2}{Are buggy lines less ``natural" than bug-fix lines?}

Even if defective lines are indeed more often fingered as unnatural by language models, it is likely to be an unreliable indication; thus one can expect many false positives (correct lines indicated as unnatural) and false negatives (buggy lines indicated as natural). It would be interesting to know, however, how well naturalness (\ie entropy) is a good ordering principle for directing inspection. 

\RQ{rq3}{Is ``naturalness" a good way to direct inspection effort?}

One can view ordering lines of code for inspection by `naturalness' as a sort of defect-prediction technique; we are inspecting lines in a certain order, because prior experience suggests that certain code is very improbable, and thus possibly defective. Traditional defect-prediction techniques typically rely on historical process data (\eg number of authors, previous changes and bugs); however, defectiveness is predicted at the granularity of files (or methods), thus, it is reasonable to compare naturalness as an ordering principle with \sta, which provide warnings at the line level.

\RQ{rq4}{How do \sta and \nbf compare in terms of ability to direct inspection effort?}

It is reasonable to expect that, if \sta provides a warning on a line \emph{and} it appears unnatural to a language model, then it is even more likely a mistake. We therefore investigate whether naturalness is a good ordering for warnings provided by static bug-finders. 

\RQ{rq5}{Is ``naturalness" a useful way to focus the inspection effort on warnings produced by \sta?}


\section{Methodology}
\label{sec:method}

In this section, we describe the projects that we studied and our approaches to data gathering and analysis.

\subsection{Study Subject}
\label{sec:study}

We studied 10 OSS java projects, as shown in Table~\ref{tab:study}: Among them \atm 
(an asynchronous web socket framework), 
\fdk (an Android SDK for building Facebook application), 
\el (a distributed search engine for cloud),
\netty (an asynchronous network application framework), and \prs (a distributed SQL query engine) 
are \gh projects, while \drb (a relational database), \luc (a text search engine library), 
\oj (a Java Persistence API), \qpd (a messaging system), and \wck (a light\-weight web application framework) 
 are taken from Apache Software Foundation. We deliberately chose the projects from different application domains
  to measure \nbf's performance in various types of systems. 
 The Apache projects are relatively older; \luc, the oldest one, started in 2001. 
The earliest \gh project in our dataset (\netty) started in 2008. 
All projects are under active development. 

We analyzed \nbf's performance on this data set in two settings. In the first setting 
(see \emph{Phase-I} in \ref{sec:dc})
we consider all the bugs---both 
development time and post release---that have appeared in the project's evolution. The performance is analyzed at different stages 
of each project's evolutionary history. We extracted snapshots of individual projects at an interval of 3 months from the version history. 
Such snapshots represent the current states of the projects at that time period (see Section~\ref{sec:dc} for details).
In total, we analyzed 266 snapshots across 10 projects that include 782,661 distinct file versions, and 202.6 Million 
total non-commented source code lines (NCSL). These snapshots contain 38,003 distinct commits, of which 8,296 were 
marked as bug fixing changes using the procedures outlined in \ref{sec:dc}. The corresponding bugs include both development-time bugs as well as post-release bugs.

\begin{table}
\scriptsize
\centering
\begin{tabular}{l|rrrrr}

 \multirow{2}{*}{Project}     &   {NCSL}     & \multicolumn{2}{c}{\#Warnings}     &  \\ 
&  \multicolumn{1}{c}{\#K} & \multicolumn{1}{c}{FindBug} & \multicolumn{1}{c}{PMD} & \multicolumn{1}{c}{\#Issues}  \\ 
\midrule
Derby (7) & 420--630 & 1527--1688 & 140--192K & 89--147 \\
Lucene (7) & \ 68--178 & 137--300 & 12--31K & 24--83   \\
OpenJPA (7) & 152--454 & 51--340 & 62--171K & 36--104  \\ 
Qpid (5)  & 212--342 & 32--66  & 69--80K & 74--127 \\
Wicket (4) & 138--178 & 45--86 & 23--30K & 47--194  \\

\end{tabular}
\caption{{\small{\bf Summary data of projects that are analyzed for locating bugs reported in issue database. The dataset is taken from 
Rahman \etal
}}}

\label{tab:study-issue}
\end{table}

In the second setting, we only focus on post-release bugs that are reported in an issue tracking system. We used the data set prepared by Rahman \etal~\cite{rahman2014comparing}, in which snapshots of the five Apache projects were taken at selected project releases. At each snapshot, the project size varies between 68 and 630K NCSL. The bugs were extracted from Apache's JIRA issue tracking system and the total number of bugs reported against each release across all the projects varies from 24-194. Table~\ref{tab:study-issue} summarizes this dataset.

At each release version, Rahman~\etal further collected warnings produced by two static bug finding tools, namely 
\fbg~\cite{ayewah2008using} and \pmd~\cite{copeland2005pmd}. \pmd operates on source code and produces line-level 
warnings; \fbg operates on Java bytecode~\cite{ayewah2008using} and reports warning at line, method, 
and class level. For this reason \fbg produces warnings covering significantly more lines, though the number of unique 
warnings is smaller than that of \pmd (see Table~\ref{tab:study-issue}). 
To make the comparisons between \nbf and \sta fair,
we further filtered out warnings for commented lines, 
since \nbf 's entropy calculation does not consider the commented lines.
In fact, we have noticed that a majority of \fbg line-level warnings are actually 
commented lines. Thus after removing comments, we are left with primarily method and class level \fbg warnings.

\subsection{Data Collection}
\label{sec:dc}

As mentioned earlier, our experiment has two distinct phases. First, we describe the process of collecting data for Phase-\rom{1},
which tries to locate all the bugs that developers fix during an ongoing development process. 
Next, we briefly summarize data collection of Phase-\rom{2},
which locates bugs at project release time;
this data set is taken from Rahman \etal~\cite{rahman2014comparing}.

\begin{figure}[!htpb]
\centering
  \includegraphics[width=0.95\columnwidth]{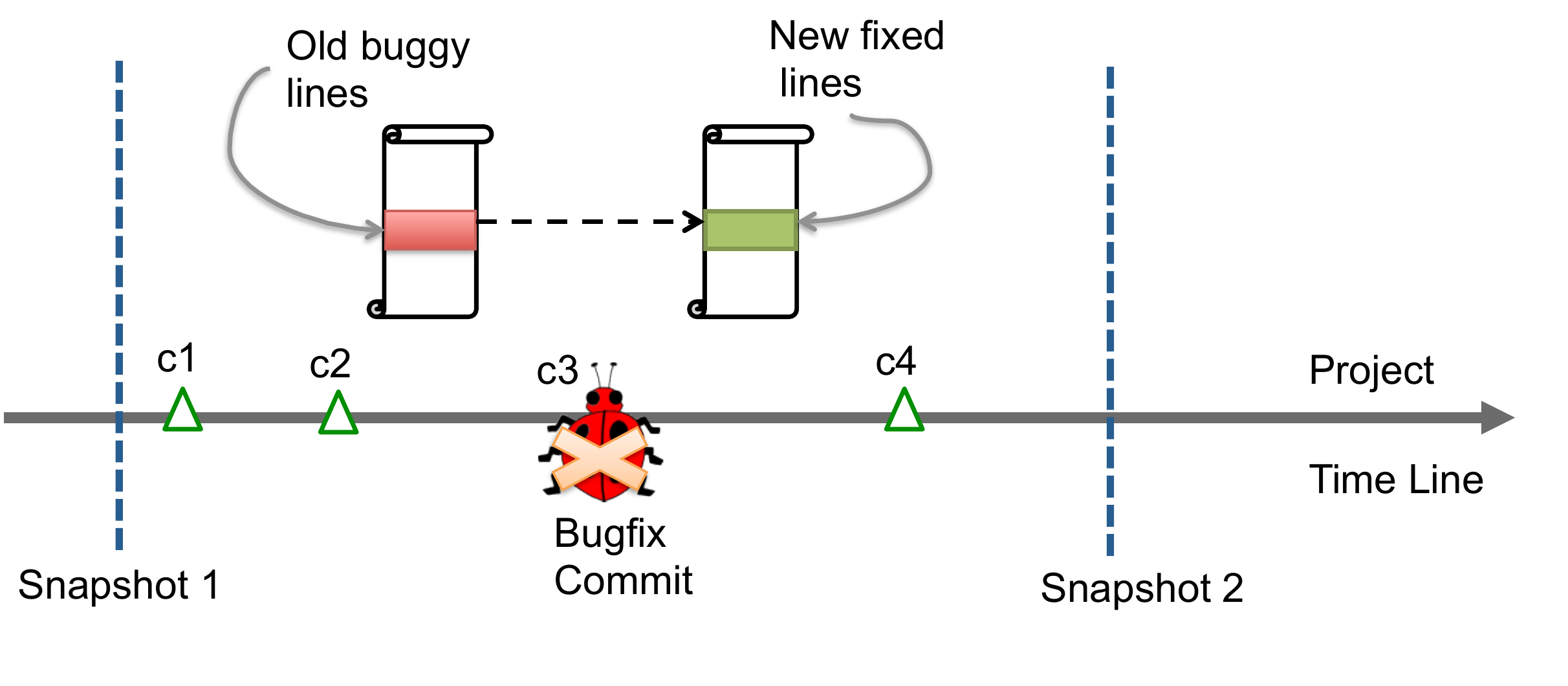}
  \caption{{\small{\bf 
Phase \rom{1} Data Collection: note Project Time Line, showing snapshots (vertical lines) and
commits (triangles) at c1$\ldots$c4. For every bugfix file commit (c3) we collect the buggy version
and the fixed version, and use {\small\tt diff} to identify buggy \& fixed lines.\\
%
%
  }}}
\label{fig:evolution}
\end{figure}

\begin{figure}[!htpb]
\centering
  \includegraphics[width=0.8\columnwidth]{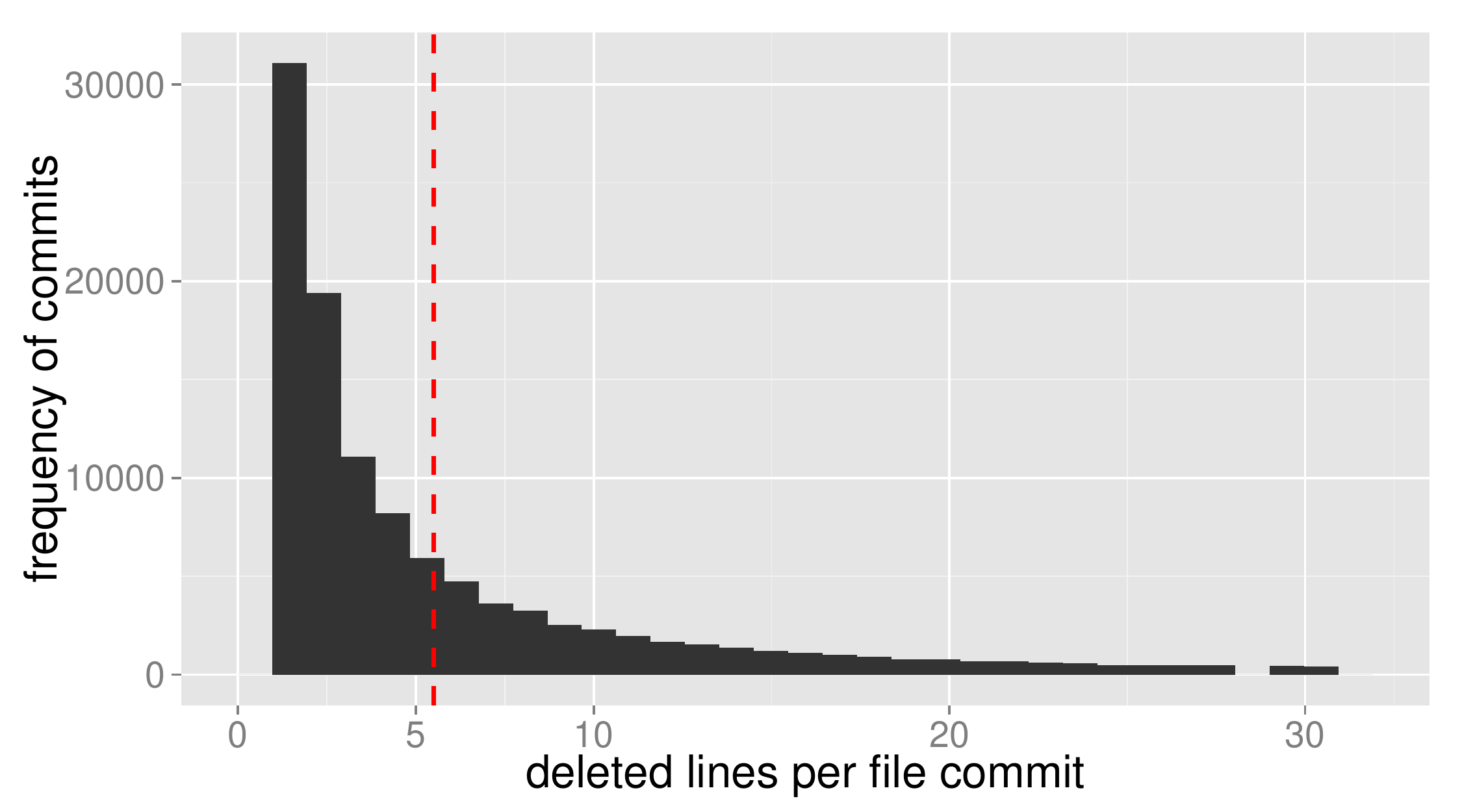}
  \caption{\textbf{\small Histogram of the number of lines deleted per file commit. The mean is 5, marked by the dashed line}
  }
\label{fig:change-size}
\end{figure}

\noindent\textbf{Phase-\rom{1}.}
All our projects  used  {\sc Git};  
we downloaded a snapshot of each project at 3-month intervals,
beginning with the project's inception.  A snapshot represents the project's state at that point of time, 
 as shown by the dashed vertical lines in Figure~\ref{fig:evolution}. 
 Then we retrieve all the commits 
 (c$_i$s in the Figure) made between each pair of consecutive snapshots. 
Each commit involves an old version of a file and its new version. Using {\tt git diff}
we identify the lines that are changed between the old and new versions. We also collected 
the number of deleted and added lines in every commit. We then removed the commits comprising 
more than 30 deleted lines (30 lines was at the 3$^{rd}$ quartile of the sample 
of deleted lines per commit in our data set). 
We further removed the commits with no deleted lines, because we are only interested in 
locating buggy lines present in the old versions. \ref{fig:change-size} shows a histogram 
of number of deleted lines per file commit; it ranges from at least one line to at max 30 lines deletion per commit 
with a median at 5.\footnote{In the unfiltered data set the median was at 2}

Each commit  has an  associated  commit log. We mark a commit as {\em bugfix}, 
if the corresponding commit log  contains at least one of the error related keywords: 
`error', `bug', `fix', `issue', `mistake', `incorrect', `fault', `defect' and `flaw', as  proposed by
 Mockus and Votta~\cite{Mockus2000}. In this step, we first convert each commit message to a 
 bag-of-words; we then remove words that appear only once among all of the bug fix messages to
  reduce project specific keywords; finally, we stem the bag-of-words using standard natural 
  language processing (NLP) techniques.  This method was taken from our previous work~\cite{ray2014lang}.
  The deleted lines corresponding to the old version of a bugfix commit are marked as {\em buggy} lines. 
The added lines associated with new corrected version are marked as {\em fixed} lines. 

Thus, from three sets of files (the files that did not change between two snapshots and the old and new versions 
of the changed files), we retrieve three sets of lines:  
(1) {\em unchanged lines}: all lines of the unchanged files and unchanged lines of the changed files. 
(2) {\em buggy lines}:  lines that were corrected in the old version of the bugfix commits, and 
(3) {\em fixed lines}: lines that were fixed in the new version of the bugfix commits. 
In total, we compared 58,374,475 unchanged lines, 88,058 buggy lines, and 204,242  fixed lines 
 across all the snapshots of all the projects.\\

\noindent\textbf{Phase-\rom{2}.} To begin with, certain release versions of each Apache project were selected. 
Then, from the JIRA issue tracking system of the Apache projects, the post-release bugfix commits 
(corresponding to the selected release) were identified. Next, by blaming the old buggy file version associated 
with a bugfix commit using {\tt git blame}, the corresponding buggy lines were detected.  Since in this phase 
we are interested in locating post-release bugs, the identified buggy lines were further mapped to the release 
time file version using an adopted version of SZZ algorithm~\cite{sliwerski2005changes}. For each project 
release version, final outcome of Phase-\rom{2} is two sets of lines: (1) {\em buggy lines}:  lines that were marked 
as buggy lines based on post release fix and (2) {\em non-buggy lines}: all the other lines across all the Java files present 
at the release version.

\subsection{Measuring entropy using cache language model}
\label{subsec:building}

\noindent\textbf{Entropy of code snippets.}
We measure the {\em naturalness} of a code snippet using statistical language model with a widely-used metric -- {\em cross-entropy} ({\em entropy} in short)~\cite{Hindle:2012:ICSE,Allamanis:2014:FSE}. The key intuition is that snippets that are more like the training corpus (\ie more natural) would be assigned higher probabilities or lower entropy from an LM trained on the same corpus.
Given a snippet $S=t_i\dots t_N$, of length $N$, with a probability $P_M(S)$ estimated by a language model $M$. The entropy of the snippet is calculated as:

\begin{equation}
	H_M(S) = -\frac{1}{N}\log_2 P_M(S) = -\frac{1}{N} \sum_{1}^{N}{\log_2 P(t_i | h)}
\end{equation}

$P(t_i | h)$ is calculated by the cache language model via Equation~\ref{equation-final-score}.\\

\noindent\textbf{Building a Cache Language Model.}
For each project and each pair of snapshots, we are interested in the entropy of lines that were marked as buggy in some commit between these snapshots. We would like to contrast these entropy scores with those of lines that were not changed in any bug-fix commit in this same period. To compute these entropy scores, for each file, we first train a language model on the `old' version of all other files (the version at the time of the previous snapshot), counting the sequences of tokens of various lengths; we then run the language model on the the current file, computing the entropy of each token based on both the prolog (the preceding tokens in the current file) and epilog (the succeeding tokens); finally, we compute the entropy of each line as the average of the entropy of each token on that line.

As an optimization step, we divided all `old' versions of files into ten bins. Then, whenever testing on a file, we use the pre-counted training set on the nine bins that the old version of the current file is not in. This removes the need to compute a training set for each file separately. Since we use the cache-based language model, the entropy scores within each file are calculated using both the training set on the other nine bins and a locally estimated cache, built on only the current file, since Tu et al.~\cite{Tu:2014:FSE} reported that building cache on both the prolog and epilog achieves best performance.\\

\begin{figure}[!htpb]
\centering
   \includegraphics[width=\columnwidth]{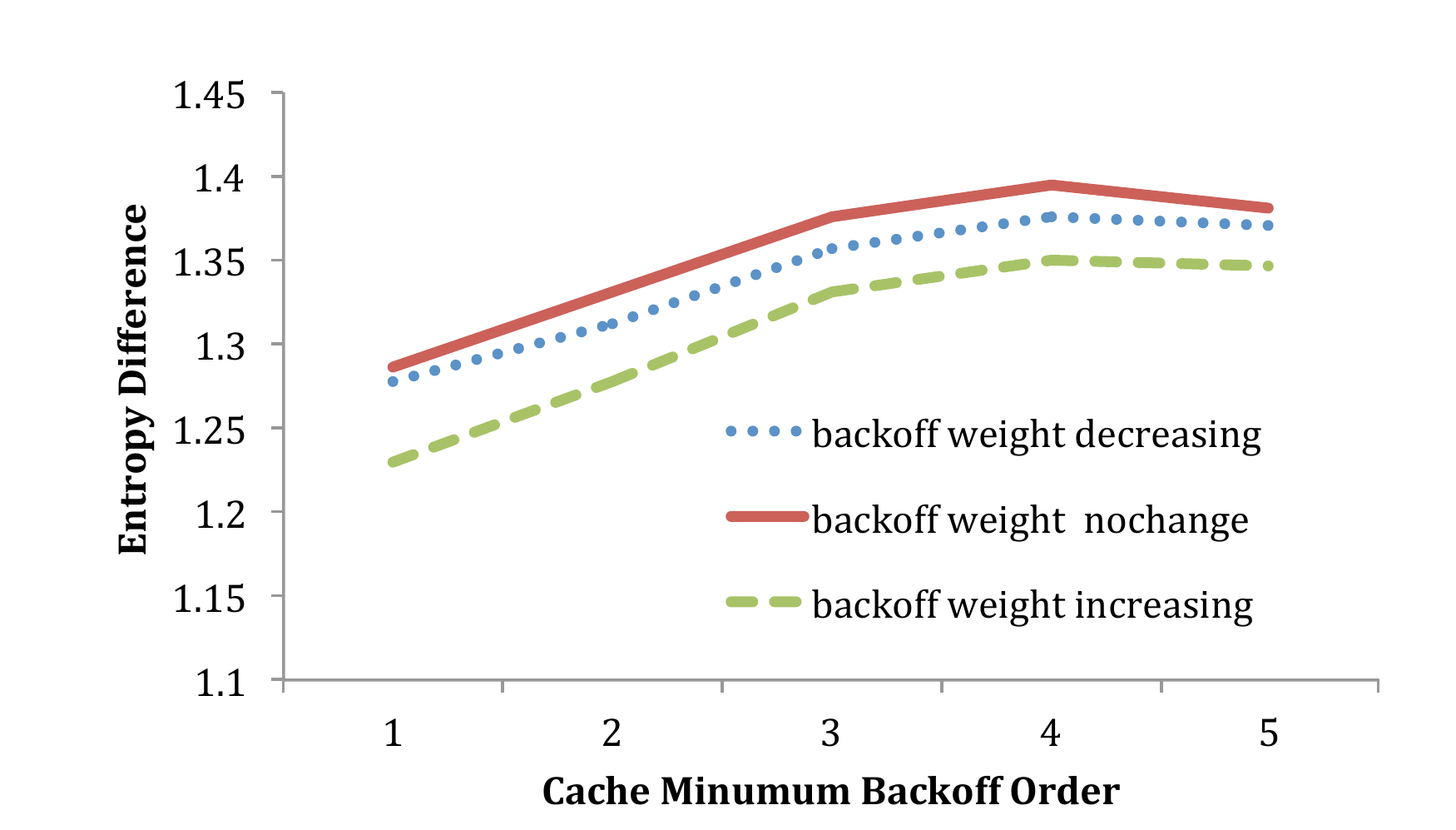}
  \caption{{\small{\bf Determining parameters of cache model. The experiments were conducted 
  on \el and \netty projects for one-line bugfix changes. Y axis represents difference of entropy of a 
  buggy line \wrt non-buggy lines in the same file. 
    }}}
\label{fig:cache-order}
\end{figure}

\vspace{1em}

\noindent\textbf{Determining parameters for cache language model.}
Several factors of the locality would affect the performance of cache language model~\cite{Tu:2014:FSE}: {\em cache context}, {\em cache scope}, {\em cache size}, and {\em cache order}.
For the fault localization task, we build the cache on all the existing code in the current file. 
In this light, we only need to tune the cache order (\ie the {\em maximum} and {\em minimum} order of \ngrams stored in the cache). In general, longer  \ngrams are more reliable but quite rare, thus we back-off to shorter matching prefixes~\cite{Katz:1987:ASSP} when needed.
We follow Tu et al.~\cite{Tu:2014:FSE} to set the maximum order of cache \ngrams to $10$. To determine the minimum back-off order, we performed experiments on the \el and \netty projects to find the optimal performance, measured in terms of difference in entropy between buggy and non-buggy lines (\ref{fig:cache-order}). The figure shows entropy difference with varying minimum backoff order and three different backoff weights (increasing, decreasing, no-change). We observed maximum difference in entropy between buggy and non-buggy lines at minimum backoff order of 4 with no change in the backoff weight.  Thus, we set the minimum backoff order be $4$ and the backoff weight be $1.0$.

\subsection{Adjusting the entropy scores}
\label{subsec:adjusting}
An important assumption underlying the applicability of language models to defect \emph{prediction} is that
higher entropy is associated with bug prone-ness. 
In practice, buggy lines are quite rare, thus a few non-buggy lines with high entropy scores could substantially increase false negatives and worsen performance. 
We undertook some tuning efforts to sharpen \nbf's prediction ability. 
\begin{table*}
\scriptsize
  \centering
    \begin{tabular}{l|rr|rr|rr|rr|rr}
    \multirow{2}{*}{project} & \multicolumn{2}{c|}{max delete = 2} & \multicolumn{2}{c|}{max delete = 5} & \multicolumn{2}{c|}{max delete = 10} & \multicolumn{2}{c|}{max delete = 20} & \multicolumn{2}{c}{max delete = 30} \\
          & \multicolumn{1}{c}{difference} & \multicolumn{1}{c|}{effect} & \multicolumn{1}{c}{difference} & \multicolumn{1}{c|}{effect} & \multicolumn{1}{c}{difference} & \multicolumn{1}{c|}{effect}
           & \multicolumn{1}{c}{difference} & \multicolumn{1}{c|}{effect} & \multicolumn{1}{c}{difference} & \multicolumn{1}{c}{effect} \\ \midrule
    atmosphere & 1.17 to 1.69 & 0.50  & 0.94 to 1.23 & 0.38  & 0.74 to 0.94 & 0.30  & 0.36 to 0.53 & 0.16  & 0.38 to 0.53 & 0.16 \\
    derby & 1.56 to 1.96 & 0.56  & 1.63 to 1.85 & 0.55  & 1.32 to 1.47 & 0.44  & 1.01 to 1.13 & 0.34  & 0.84 to 0.94 & 0.28 \\
    elasticsearch & 1.58 to 1.90 & 0.57  & 1.37 to 1.53 & 0.48  & 1.01 to 1.14 & 0.35  & 0.84 to 0.95 & 0.29  & 0.73 to 0.82 & 0.25 \\
    fdk & 0.74 to 1.64 & 0.40  & 1.33 to 1.83 & 0.53  & 1.03 to 1.41 & 0.41  & 1.04 to 1.33 & 0.40  & 0.92 to 1.17 & 0.35 \\
    lucene & 1.27 to 1.80 & 0.48  & 0.97 to 1.28 & 0.35  & 0.83 to 1.06 & 0.30  & 0.97 to 1.14 & 0.33  & 0.79 to 0.96 & 0.28 \\
    netty & 1.97 to 2.24 & 0.68  & 1.58 to 1.74 & 0.54  & 1.32 to 1.44 & 0.45  & 1.12 to 1.22 & 0.38  & 0.98 to 1.07 & 0.33 \\
    openjpa & 1.61 to 2.12 & 0.59  & 1.15 to 1.42 & 0.41  & 0.89 to 1.10 & 0.32  & 0.68 to 0.84 & 0.24  & 0.60 to 0.75 & 0.21 \\
    presto & 1.12 to 1.73 & 0.47  & 0.95 to 1.30 & 0.37  & 0.88 to 1.14 & 0.33  & 0.76 to 0.96 & 0.28  & 0.72 to 0.90 & 0.27 \\
    qpid  & 1.35 to 1.75 & 0.51  & 1.19 to 1.40 & 0.42  & 0.98 to 1.13 & 0.35  & 0.65 to 0.77 & 0.23  & 0.58 to 0.68 & 0.21 \\
    wicket & 1.51 to 1.88 & 0.56  & 1.44 to 1.64 & 0.51  & 1.18 to 1.33 & 0.41  & 0.95 to 1.08 & 0.33  & 0.92 to 1.03 & 0.32 \\
    \midrule
    overall & 1.67 to 1.80 & 0.56 & 1.41 to 1.48 & 0.47 & 1.13 to 1.18 & 0.37 & 0.91 to 0.95 &  0.30 & 0.80 to 0.84 & 0.26 \\ 
    \end{tabular}%
  \caption{\textbf{\small Buggy lines, in general, have higher entropy than non-buggy lines. 
    Difference is measured with t-test for 95\% confidence interval, and effect is
      Cohen's D. Wilcox non-parametric test also confirmed buggy lines have higher entropy with statistical significance.
     `max delete' represents maximum number of buggy lines that are fixed in a file commit. 
    }}
  \label{tab:ld}%
\end{table*}%

We manually examined entropy scores of sample lines and found strong associations with lexical and syntactic properties. 
In particular, lines with many and/or previously unseen identifiers, such as package, class and method declarations, had substantially higher entropy scores than average. Lines such as the first line of for-statements and catch clauses had much lower entropy scores, being often repetitive and making use of earlier declared variables. We use these variations in entropy scores by introducing the notion of \emph{line types}, based on the code's syntactic structure, \ie the abstract syntax tree (AST), and computed a syntax-sensitive entropy score. 

First, with each line, we associated syntax-type, corresponding to the grammatic entity that is the lowest  AST node that includes the full line.
These are typically  AST node types such as statements (\eg if, for, while), 
declarations (\eg variable, structure, method) or nodes that typically span one line, such as switch cases and annotations.
We  then compute a normalized Z-score for the entropy of the line, \emph{over all lines with that node type}. 

\begin{equation*}
z_\text{line, type} = \frac{entropy_\text{line} - \mu_\text{type}}{SD_\text{type}}
\end{equation*}

The above normalization essentially uses 
 the extent to which a line is ``unnatural'' with respect to other lines of the same type, 
 to  predict how likely it is to be  buggy. 
 In addition, we don't expect all line types to be equally buggy; package declarations ({\small\tt import$\ldots$}) are probably usually correct,
when compared to  error handling \\({\small\tt try$\ldots$catch}).
The previously computed line-types come in handy here too: we can compute the \emph{relative bug-proneness} of a type based on the fraction of  bugs and total lines it had in all previous snapshots. Hence, we use the first snapshot as a `training set' for this model and compute the bug-weight of a statement as:

\begin{equation*}
w_\text{type} = \frac{bugs_\text{type}/lines_\text{type}}{\sum_\text{t$\in$types} bugs_\text{t}/lines_\text{t}}
\end{equation*}

where the bugs and lines of each type are counted over all previous snapshots. We then scale the z-score of each line by it's weight $w$ to achieve our final model, which we name \ngmwt.\\


\section{Evaluation}
\label{sec:eval}
\setcounter{RQCounter}{0}
\noindent We begin with the question at the core of this paper: 

\RQA{rq1a}{Are buggy lines different from non-buggy lines?}

For each project, we compare line entropies of buggy and non-buggy lines. For a given
snapshot,  non-buggy lines consist of all the unchanged lines and the deleted lines that are
not part of a bug-fix commit. The buggy lines include all the deleted lines in all bug-fix
commits. Figure~\ref{fig:rq1} shows the result, averaged over all the studied projects.
Buggy lines are associated with higher entropies. Table~\ref{tab:ld} further details the average
entropy difference between the buggy and non-buggy lines (buggy $>$ non-buggy) and the effect
sizes (Cohen's D) between the two.  Wilcox non-parametric test confirms the difference with
statistical significance (p-value $<$ 2.2*10$^{-16}$).

Note that both entropy difference and effect size decrease as we increase the threshold for the maximum number
of deleted lines (max\_delete) in a file commit.  For example, for a max\_delete size of 2, the entropies of
buggy lines are on average 1.67 to 1.80 bits higher, with a high effect size of
0.56. However, when we consider all the studied commits (max\_delete = 30), the entropies of buggy lines are, on average, 
less than a bit (0.80 to 0.84 bits) higher, with a small-to-moderate effect size
of 0.26. One possible explanation: particularly in larger bug-fix commits, some of the 
deleted (or modified) lines might only be indirectly associated with the erroneous lines
that most improbable (``unnatural''). These indirectly associated lines might actually be common, and thus
have lower entropy; this would diminish overall entropy
differences between the buggy and non-buggy lines. However, with statistical significance, we have the overall result: 

\begin{figure}[!htpb]
\centering
\includegraphics[width=\columnwidth]{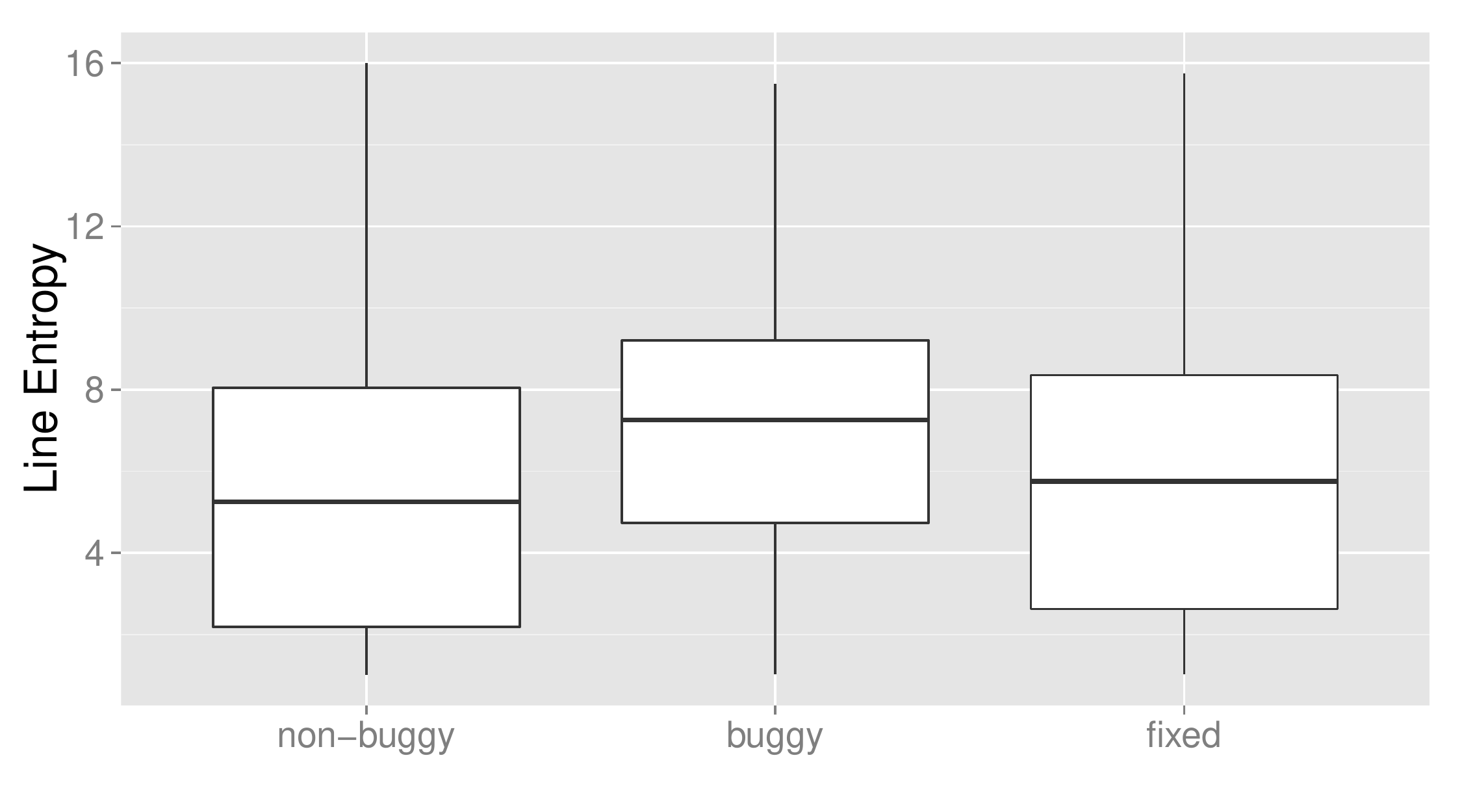}
   \caption{\textbf{\small Entropy difference between non-buggy, buggy, and fixed lines. 
  File commits upto 5 deleted lines are considered, since five is the average number of deleted lines per file commit
  (see Figure~\ref{fig:change-size}). }}
\label{fig:rq1}
\end{figure}
 
\RS{rq1a}{Buggy lines, on average, have higher entropies than non-buggy lines.}

{\samepage
\begin{table}
\centering
\begin{tabular}{p{0.45\textwidth}}
\toprule
\small{\bf {Example 1} :  Wrong Initialization Value} \\ 
{\small Facebook-Android-SDK (2012-11-20)} \\
{\small File: ~\texttt{Session.java}} \\
\small{Entropy dropped after bugfix : {\bf 4.12028} } \\
\begin{lstlisting}[ language=java]
  if (newState.isClosed()) {
    // Before (entropy = 6.07042):
-   this.tokenInfo = @\Red{null}@;
    // After (entropy = 1.95014):
+   this.tokenInfo = @\dkgreen{AccessToken.createEmptyToken}@
                       @\dkgreen{(Collections.<String>emptyList())}@;
  }
  ...
\end{lstlisting} \\ \\

\midrule
\small{\bf {Example 2} :  Wrong Method Call} \\ 
{Netty (2013-08-20)} \\
File: ~\texttt{ThreadPerChannelEventLoopGroup.java} \\ \\
\small{Entropy dropped after bugfix : {\bf 4.6257} }\\

\begin{lstlisting}[ language=java]
  if (isTerminated()) {
    // Before (entropy = 5.96485):
-   terminationFuture.@\Red{setSuccess}@(null);
    // After (entropy = 1.33915):
+   terminationFuture.@\dkgreen{trySuccess}@(null);
   }
\end{lstlisting} \\

\midrule
\small{\bf {Example 3} :  Unhandled  Exception} \\ 
{\small Lucene (2002-03-15)} \\
{\small File: ~\texttt{FSDirectory.java}} \\
\small{Entropy dropped after bugfix : {\bf 3.87426} } \\

\begin{lstlisting}[ language=java]
  if (!directory.exists())
    // Before (entropy = 9.213675):
-   @\Red{directory.mkdir();}@
    // After (entropy = 5.33941):
+   @\dkgreen{if (!directory.mkdir())}@
+     @\dkgreen{throw new IOException}@
           @\dkgreen{("Cannot create directory: " + directory)}@;

...
\end{lstlisting} \\ \\

\bottomrule
\end{tabular}
\caption{\small{\bf Examples of bug fix commits that \nbf detected successfully. 
These bugs evinced a large entropy drop after the fix.
Bugs with only one defective line are shown for simplicity purpose.
The errors are marked in~\red{red}, and the fixes are highlighted in~\dkgreen{green}.
}}
\label{sampletab}
\end{table}

\begin{table}
\centering
\begin{tabular}{p{0.45\textwidth}}
\toprule
{\bf {Example 4} :  Wrong Argument (\nbf could not detect)} \\ 
{Netty (2010-08-26)} \\
File: ~\texttt{HttpMessageDecoder.java} \\ 
Entropy increased after bugfix : {\bf 5.75103}  \\

\begin{lstlisting}[ language=java]
if (maxHeaderSize <= 0) {
   throw new IllegalArgumentException(
     // Before (entropy = 2.696275):
-    "maxHeaderSize must be a positive integer: "
                          + @\Red{maxChunkSize)}@;
     // After (entropy = 8.447305):
+    "maxHeaderSize must be a positive integer: "
                          + @\dkgreen{maxHeaderSize)}@;
}
\end{lstlisting} \\ \\

\toprule

{\bf {Example 5} :  (\nbf detected incorrectly)} \\
{Facebook-Android-SDK (multiple snapshots)} \\
File: ~\texttt{Request.java} \\ 

\begin{lstlisting}[ language=java]
   // Entropy = 9.892635
   Logger logger = new Logger(LoggingBehaviors.REQUESTS, "Request");
    ...
\end{lstlisting} \\ \\

\bottomrule
\end{tabular}
\caption{\small{\bf Examples of bug fix commits where \nbf did not perform well. In Example 4, \nbf could not detect the bug 
successfully (marked in~\red{red}) and after bugfix the entropy has increased. In Example 5, \nbf incorrectly detected the line as buggy due to its high entropy value.}} 
\label{tab:sample}
\end{table}

}
\smallskip


\RQA{rq2a}{Does entropy of a buggy line drop after the bug is fixed?}
In a bug-fix commit, the lines deleted from the original versions are considered 
\emph{buggy lines} and lines added in fixed versions are considered  \emph{fixed lines}. To
answer RQ2, we collected all the buggy and the fixed lines across all the projects and compared
their average entropies.  It is hard to establish a one-to-one correspondence
between a buggy and a fixed line, because often buggy lines are fixed by a different number of new lines.
Hence, we compare the mean entropies between buggy and non-buggy hunks. \ref{fig:rq1} shows the result. On average, entropy of the buggy lines drop after the bug-fixes, with a drop of 1.19 to 1.26 bit, with 95\% confidence. 

~\ref{sampletab} shows three examples of code where entropy of buggy lines dropped
significantly after bug-fixes. In the first example, a bug was introduced in {\tt
Facebook-Android-SDK} code due to a wrong initialization value\textemdash {\tt tokenInfo} was
incorrectly reset to null (see the commit log).  This specific initialization rarely occurred elsewhere, 
so the buggy line had a rather high entropy of $6.07$. Once the bug was fixed,
the fixed line followed a repetitive pattern (indeed, with two prior instances in the same
file). Hence, entropy of the fixed line dropped to $1.95$, an overall 4.12 bit reduction.  The
second example shows an example of incorrect method call in the {\tt Netty} project. Instead of
calling the method {\tt trySuccess} (used three times earlier in the same file), the code 
incorrectly called the method {\tt setSuccess}, which was never called in a similar context. After
the fix entropy drops by 4.6257 bits. Finally, example 3 shows an instance of missing
conditional check in {\tt Lucene}. The developer should check whether directory creation is
successful by checking return value of {\tt directory.mkdir()} call, following the usual code
pattern. The absence of this check raised the entropy of the buggy line to $9.21$. The
entropy value drops to $5.34$ after the fix.

The table below shows the average drop of entropy and the Cohen's D effect size of buggy vs.~ fixed
lines, with varying thresholds for the maximum bug size in terms of deleted lines. 
\begin{center}
{
  \scriptsize
    \begin{tabular}{llllll}

    \multicolumn{1}{l}{Max Delete} & 2     & 5     & 10    & 20    & 30 \\
    \midrule
    \multicolumn{1}{c}{\multirow{2}[0]{*}{\begin{tabular}[x]{@{}c@{}}Entropy drop \\ ~after bugfix\end{tabular}}} & 1.52  to & 1.19 to & 0.88 to  & 0.65 to & 0.53 to  \\
    \multicolumn{1}{c}{} & 1.62  & 1.26  & 0.94  & 0.70  & 0.58 \\
    \multicolumn{1}{l}{Effect Size} & 0.51  & 0.40  & 0.30  & 0.22  & 0.18 \\

    \end{tabular}%
}
\end{center}

Similar to the result of RQ1, both the entropy
difference and effect size vary with maximum delete threshold: when the delete threshold increases, the other two decrease.
For example, at maximum delete size 2, mean entropy drops from 1.52 to
1.62 bit (with 95\% confidence) with statistical significance, with a large effect size ($>$
$0.50$). However, with delete threshold at 30,  mean entropy difference between the buggy and
non-buggy lines are only half a bit with a small effect size of 0.18. For all the studied
ranges, the Wilcox non-parametric test confirms with statistical significance that the entropy of buggy lines is higher than the entropy of the fixed lines.

However, in certain cases these observations do not hold. For instance, in the example 4 of Table~\ref{tab:sample}, entropy increased after the bug fix by 5.75 bits. In this case, developer copied {\tt maxChunkSize} from a different context but forgot to update the variable name. This is a classic example of copy-paste error~\cite{ray2013detecting}. Since, the statement related to {\tt maxChunkSize} was already present in the existing corpus, the line was not surprising. Hence, its  entropy was low although it was a bug. When the new corrected statement with {\tt maxHeaderSize} was introduced, it increased the entropy. Similarly, in Example 5 of Table~\ref{tab:sample}, the statement related to {\tt logger} was newly introduced in the corpus. Hence, its entropy was higher although it was not a bug.

\RS{rq2a}{Entropy of the buggy lines drops after bug-fixes, with statistical significance.}

\begin{figure*}[!htpb]
\centering
\subfigure[Overall AUCEC upto inspecting 20\% lines for all the projects]{
  \includegraphics[width=0.9\columnwidth,height=5cm]{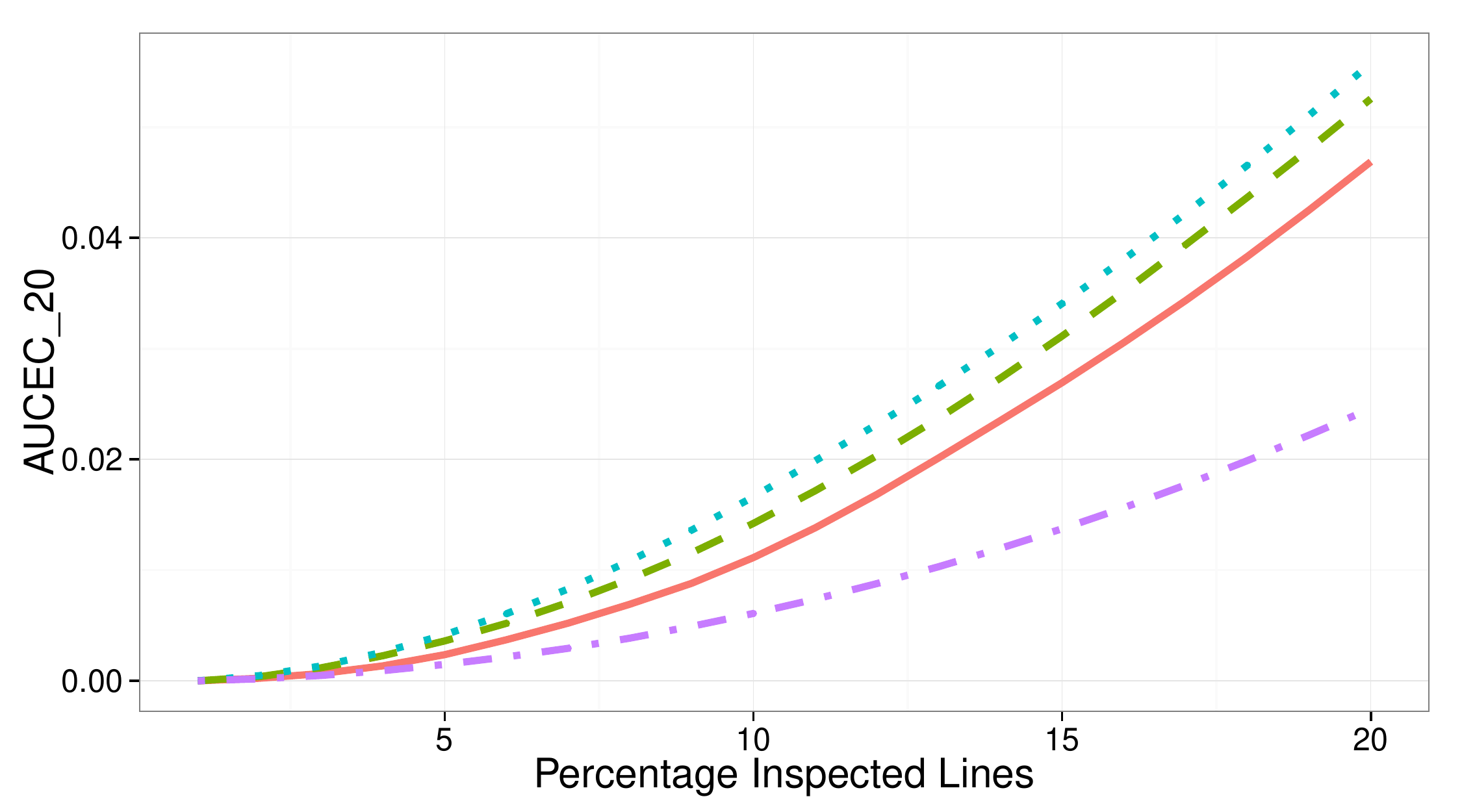}
\label{fig:cons_all}}
\quad
\subfigure[Closer look at low order AUCEC, upto inspecting 5\% lines for individual project]{
  \includegraphics[width=\columnwidth,height=5cm]{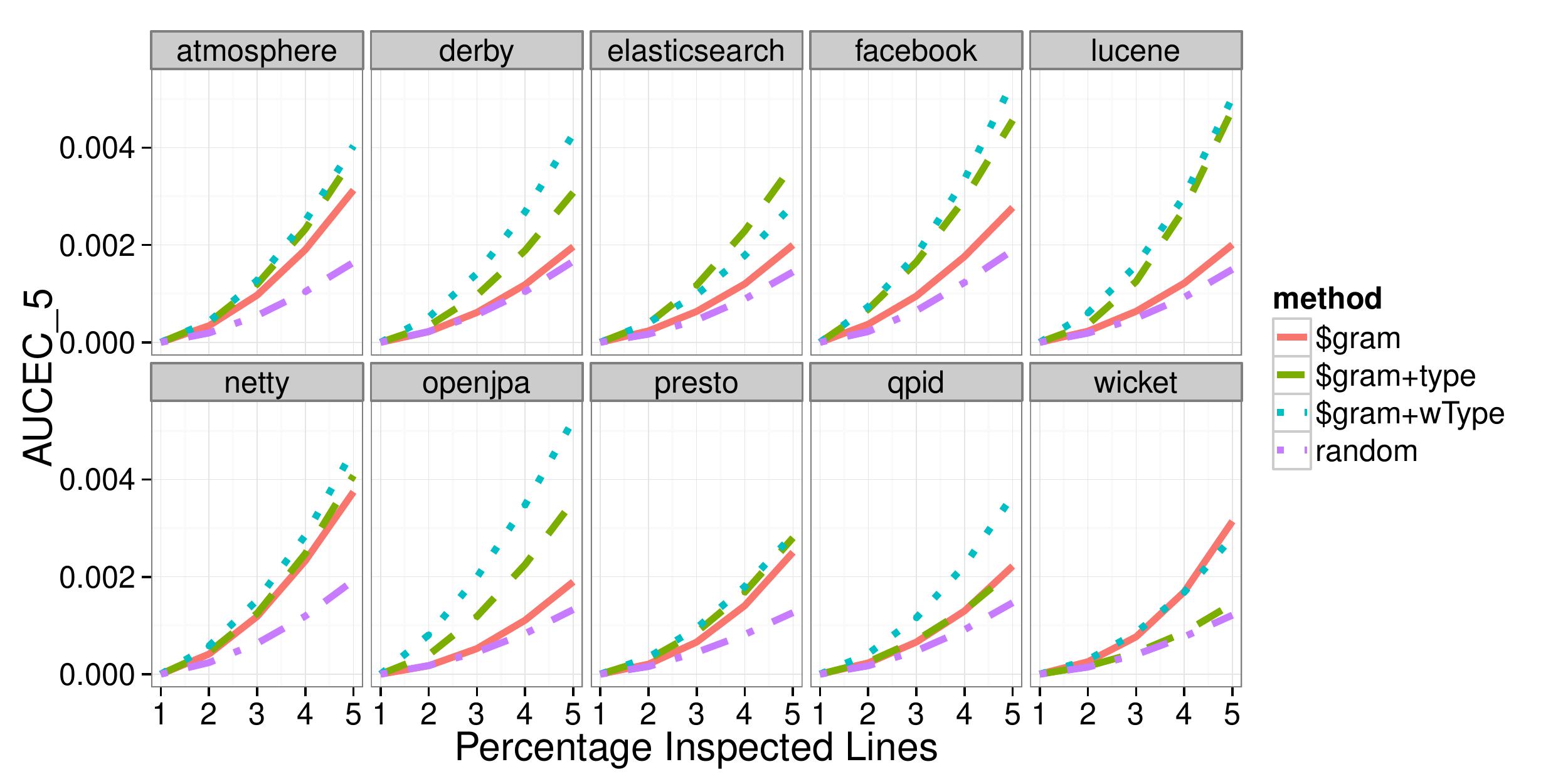}
\label{fig:conservative2}
}
\vspace{-0.1cm}
\caption{\small{\bf Performance Evaluation of \nbf with Partial Credit.}}
\label{fig:conservative}
\end{figure*}

\RQA{rq3a}{Is ``naturalness" a good way to direct inspection effort?}
Having established that buggy lines are significantly less natural than non-buggy lines, we investigate whether entropy of a line can be used 
to direct inspection effort towards buggy code. 
In particular, we start by asking whether ordering lines by entropy will better guide inspection effort that ordering lines at random. For the reasons outlined in \ref{subsec:evaluating}, we evaluate the performance of entropy-ordering, with the AUCEC scores at 5\% of inspected lines (AUCEC$_5$ in short). Furthermore, as outlined in section \ref{subsec:evaluating}, we evaluate performance according to two types of \emph{credit}: partial and full (in decreasing order of strictness). Finally, we disregarded all bugs that were part of a bug-fix which removed 15 or more lines, which we found this to be the 95th percentile of bug-fix sizes. As shown in \ref{tab:ld}, entropy plays a substantially smaller role in lines belonging to larger bug-fixes, hence we leave the identification of these lines to future research. We remind the reader that 
AUCEC$_5$ is a non-parametric, cost-sensitive measure, and the comparison to random choice is done on an equal credit basis. 

\ref{fig:cons_all} shows the AUCEC scores for partial credit, averaged over all projects, up to 20\% of the inspected lines. \ref{fig:conservative2} offers a closer look at the performance on the 10 studied projects, up to 5\% of the inspected lines. We see that, under partial credit, the default \ngm model  (without the syntax weighting described in \S\ref{subsec:adjusting})
performs significantly better than random, particularly at more than 10\% of inspected lines. However, at 5\% of inspected lines its performance varies, consistently performing better than random but often just slightly. Indeed, average performance of \ngm was significantly better than random at 20\% (nearly twice as good) but only marginally so at 5\% (17\% better than random).

This picture changes substantially with the introduction of linetypes. Scaling the entropy scores by line type improves AUCEC$_5$ performance in all but one case (Wicket) and significantly improves performance in all cases where \ngm performed no better than random. Including the bugginess history of linetypes (\ngmwt) furthermore improves prediction performance in all but one system (Elasticsearch). The latter model consistently outperforms random and \ngm (except on Wicket), achieving AUCEC$_5$ scores of more than twice that of random. These results were quite the same under full credit. Since \ngmwt is the best-performing ``naturalness" approach, we hereafter refer to it as \nbf. 

\RS{rq3a}{Entropy, is a better way to choose lines for inspection than random}

In previous work, Rahman \etal compared static bug-finders with statistical defect prediction approaches \cite{rahman2014comparing}. To this end, they created a dataset consisting of 32 releases of 5 popular Apache projects and annotated the lines in each release with both bug information and \sta information, as described in \S\ref{sec:study}. Among others, they found that ordering \sta warnings based on statistical defect prediction methods can improve the native \sta ordering. This provides an interesting challenge for the \nbf algorithm: how does the \nbf algorithm compare to \sta, and can we improve the default ordering of the \sta by using the techniques presented before? We investigate this in the next two research questions.

\begin{figure*}[!htpb]
\centering
\subfigure[AUCEC$_5$ performance of \ngmwt vs.~\pmd and the combination model]{
 \includegraphics[width=0.95\columnwidth,height=4.5cm,trim=25 0 25 0]{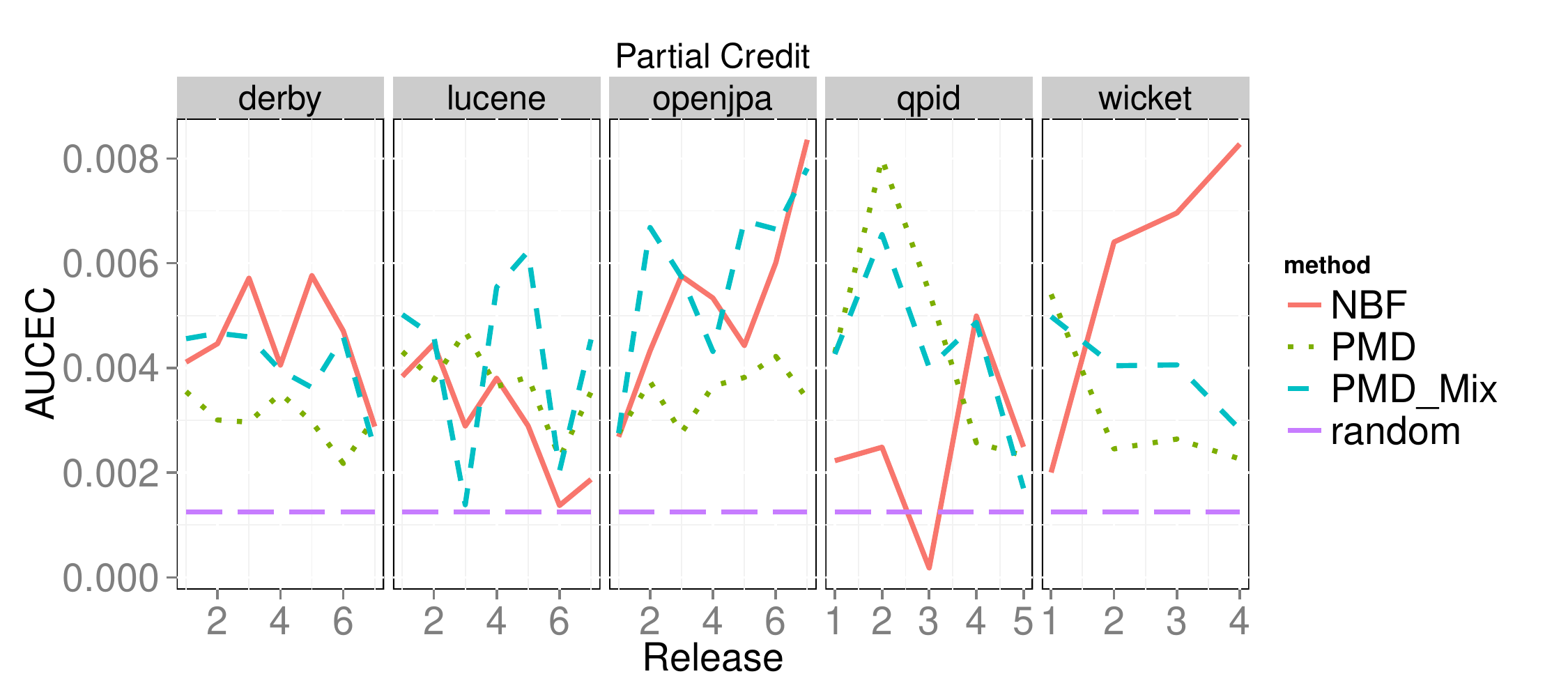}
\label{fig:ld1}}
\quad
\subfigure[AUCECL performance of \ngmwt vs.~\pmd and the combination model]{
  \includegraphics[width=0.95\columnwidth,height=4.5cm,trim=25 0 25 0]{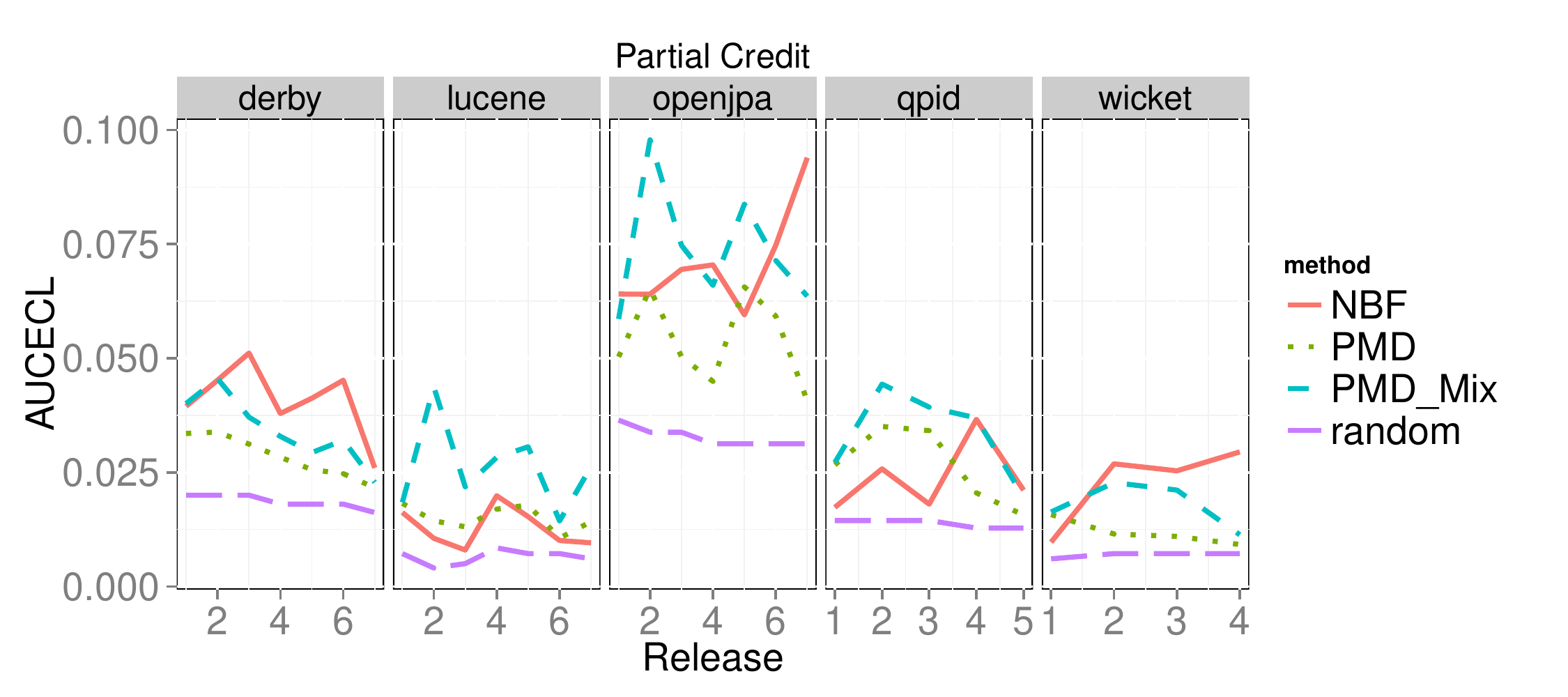}
\label{fig:ld2}
}
\vspace{-0.1cm}
\caption{\small{\bf Partial Credit: Performance Evaluation of \ngmwt \wrt \sta and a mix model which ranks the \sta lines by entropy. }}
\label{fig:sbf-vs-sta}
\end{figure*}

\RQA{rq4a}{How do \sta and \nbf compare in terms of ability to direct inspection effort?}

To compare \nbf with \sta, we computed entropy scores for each line in Rahman's dataset using the \ngmwt model. Here we again use the threshold of 14 lines for bugs, which roughly corresponded to the fourth quartile of bug-fix sizes on this dataset. Indeed, we found that Rahman's dataset had substantially more `large' bugs compared to our earlier experiments, hence we also report results without imposing this threshold.

Rahman \etal developed a measure named AUCECL to compare \sta and \dpm methods on an equal footing. In this method, the \sta under investigation sets the line budget based on the number of warnings it returns and the \dpm method may choose a (roughly) equal number of lines. The models' performance can then be compared by computing the AUCEC scores both approaches achieve on the same budget. We repeat to compare \sta with \nbf. 

Furthermore, we also compare the AUCEC$_5$ scores of the algorithms. For the \ngmwt model this is analogous to the results in RQ3. To acquire AUCEC$_5$ scores for the \sta, we simulate them as follows: First assign each line the value zero if it was not marked by the \sta and the value of the \sta priority otherwise (\{1, 2\} for \fbg, \{1 - 4\} for \pmd); then, add a small random amount (tie-breaker)
from $U[0,1]$ to all line-values and order the lines by descending value. This last step simulates the developer randomly choosing to investigate the returned by \sta: first from those marked by the \sta in descending (native, \sta tool-based) priority, and within each priority level at random. We repeat the simulation multiple times and average the performance.

\ref{fig:ld1} and \ref{fig:ld2} show the AUCEC$_5$ and AUCECL scores for \pmd on the dataset by Rahman \etal \cite{rahman2014comparing} using partial credit. The results for \fbg were comparable, as were the results using full credit. As can be seen, performance varied substantially between projects and between releases of the same project. Across all releases and under both AUCEC$_5$ and AUCECL scoring, all models performed significantly better than random (paired t-test: $p<10^{-3}$), with large effect (Cohen's D > 1). \sta and \nbf performed comparably; \nbf performed slightly better when using both partial credit and the specified threshold for bug-sizes, but when dropping the threshold, and/or with full credit, no significant difference remains between \nbf and \sta. No significant difference in performance was found between \fbg and \pmd either.

In all comparisons, all approaches retrieved relatively bug-prone lines by performing substantially better than random.

\RS{rq4a}{Entropy achieves comparable performance to commonly used \sta in defect prediction.}

Notably, \nbf had both the highest mean and standard deviation of the tested models, whereas \pmd's performance was most robust. This suggest a combination of the models: We can order the warnings of the \sta using the \ngmwt model. In particular, 
we found that the standard priority ordering of the \sta is already powerful, so we propose to re-order the lines \emph{within each priority category}. 
%
%
%

\RQA{rq5}{Is ``naturalness" a useful way to focus the inspection effort on warnings produced by \sta?}
Given the comparable performance of the \sta and \nbf models and the robustness of the \sta algorithms, we may expect a combination of the models to yield superior performance. To this end, we again assigned values to each line based on the \sta priority as in RQ4. However, rather than add random tie-breakers, we rank the lines within each priority bin by the (deterministic) \ngmwt score. The results for \pmd are shown in \ref{fig:sbf-vs-sta}, first using the AUCEC$_5$ measure (\ref{fig:ld1}) and then using the AUCECL measure (\ref{fig:ld2}). PMD\_Mix refers to the combination model as proposed.

Overall, the combined model produced the highest mean performance in both categories. It significantly outperformed the two \sta{}s  in all cases ($p<0.01$) and performed similarly to the \nbf model (significantly better on Lucene and QPid, significantly worse on Derby ($p<0.05$), all with small effect). These results extended to the other evaluation methods, using full credit and/or removing the threshold for max bug-fix size. In all cases, the mix model was either significantly better or no worse than any of the other approaches when averaged over all the studied releases.

We further evaluated ranking all warnings produced by the \sta by entropy (\emph{ignoring} the \sta priorities) and found comparable but slightly weaker results.
These results suggests that both \nbf and \sta contribute valuable information to the ordering of bug-prone lines and that their combination yields superior results.

\RS{rq5}{Ordering \sta warnings by priority \underline{and} entropy significantly improves \sta performance.}


\section{Threats to Validity}
\label{sec:threats}
A number of threats to the internal validity of the study arise from the experimental setup.\\

\noindent{\bf Identifying Buggy Lines.} The identification of buggy lines is a possible source of error. 
We used the procedure as proposed by Mockus and Votta to identify bugfix commits~\cite{Mockus2000}, which may have lead to both false negatives and false positives in the identification of buggy lines.

Programmers may fail to indicate bug fixes in log messages, leading to false negatives (missing bugs). 
There is no reason to suspect that these missing buggy lines have a significantly different entropy-profile. 
Second, as yet unfixed bugs may linger in the code, constituting a right-censorship of our data; these bugs might
have a different entropy-profile 
(although, again, no reason to suspect that this is so). 
Finally, it has been noted that developers may combine multiple unrelated changes in one commit \cite{herzig2011untangling, diasuntangling}.
This work  (\emph{op. cit.}) observed that, when multiple changes were combined, bug-fix commits were mostly combined with refactorings and code formatting efforts. This may have lead to the deletion of non-buggy high-entropy lines in a bugfix commit, although we expect these lines to form the minority of studied lines.

The threats identified above may certainly have lead to the misidentification of some lines; however,  given the high significance of the difference in entropy between buggy and non-buggy lines, we consider it unlikely that these threats could invalidate our overall results. Furthermore, the performance in defect prediction of a model using entropies on the (higher quality, JIRA-based) bug dataset by Rahman \etal confirms our expectations regarding the validity of these results. A final threat regarding RQ2 is the identification of `fixed' lines, lines that were added in the place of `buggy' lines during a bugfix commit. It is possible that the comparisons between these categories is skewed, \eg because bugfix commits typically replace buggy lines with a larger number of fixed lines. We found no evidence of such a phenomenon but acknowledge the threat nonetheless. Future research may apply entropy to defect \emph{correction} and further study this relation of buggy lines to fixed lines in terms of entropy.

In RQ4 and RQ5 we investigated the performance of the proposed \nbf in comparison to (and in combination with) static bug finders. We note that here too the identification of buggy lines may be a cause for systematic error, for which we point both to the above discussion and to Section 5 of Rahman \etal, in which they identify a number of threats to the validity of their study \cite{rahman2014comparing}.\\

Our comparison of \sta and \nbf assumes that indicated lines are equally informative to the inspector, which is not entirely fair; \nbf just marks a line as ``surprising'', whereas \sta provides specific warnings. On the other hand, we award credit to \sta whether or not the bug has anything to do with the warning on the same lines; indeed, earlier work~\cite{thung2012extent} suggests that warnings are not often related to the buggy lines which they overlap. So this may not be a major threat to our RQ4 results. 

Finally, the use of AUCEC to evaluate defect prediction has been criticized for ignoring the cost of false negatives~\cite{zhang2013cost}; the development of better, widely-accepted measures remains a topic of future research. 

\smallskip

\noindent{\bf Generalizability.} The selection of systems constitutes a potential threat to the external validity of this research. We attempted to minimize this threat by using systems from both Github and Apache, having a substantial variation in age, size and ratio of bugs to overall lines (see table \ref{tab:study}).

Finally, does this approach generalize to other languages? There's nothing language-specific about the \emph{implementation} of n-gram and \cgram models  (the \ngmwt model, however, does require parsing, which depends on language grammar). Our own prior research~\cite{Hindle:2012:ICSE,Tu:2014:FSE} showed that these models work well to capture regularities  languages such as  Java, C, and Python, yielding low cross-entropies when well trained on a corpus of code. The question remains then whether language models can identify buggy code in other languages as well, and whether the entropy drops upon repair. As a sanity check, using the same methods described in \ref{sec:method}, we gathered data from $3$ C/C++ projects ({\tt Libuv}, {\tt Bitcoin} and {\tt Libgit}). These projects together constituted just over $10$M LOC. We gathered snapshots spanning the period of November 2008 - January 2014. The data comprised a total of $8298$ commits, including $2518$ bug-fix commits (identified as described in \ref{sec:method}). We lexicalized, and parsed these projects and computed entropy scores over buggy lines, fixed lines, and non-buggy lines as described earlier. The results were fully consistent with those presented in \ref{tab:ld} and \ref{fig:rq1}; we found that buggy lines were between 0.87 and 1.16 bits more entropic than non-buggy llines when using a threshold of 15 lines (slightly smaller than among the Java projects). For a threshold of 2 lines, this difference was between 1.61 and 2.23 bits (slightly larger than in \ref{tab:ld}). Furthermore, the entropy of buggy lines (when using a threshold of 15 lines) dropped by nearly one bit on this dataset as well. These findings mitigate the external validity threat of our work, and strongly suggest that our results generalize to C/C++; we are investigating the applicability to other languages. 

\section{Related Work}
\label{sec:related}

In the following we analyze work related to our investigation.

\subsection{Statistical Defect Prediction}

Software development is an incremental process. This incremental progress is successfully logged in version control systems 
like git, svn and issue databases. Learning from such historical data of reported (and fixed)
  bugs, Statistical Defect Prediction (\DP) aims to 
 predict location of the defects that are yet to be detected. This is a very active area (see ~\cite{catal2009systematic} for a survey of the area), even having the dedicated PROMISE series of conferences (See ~\cite{Wagner:2014:2639490} for recent proceedings). 
The state of the art \DP not only leverages bug history, it also takes 
 into account several other product (file size, code complexity, code churn etc.) and process metrics~\cite{rahman2013and} (developer count, code ownership, developer experience, change frequency \etc.) to improve the prediction model. Thus, using different supervised learning techniques like logistic regression, 
 Support Vector Machine (SVM) etc., \DP associates various software entities (e.g., methods, files and packages) with their respective defect proneness.
 
  Given a fixed budget of SLOC that needs to be inspected to effectively find most bugs, \DP ranks files that one should inspect to detect most of the errors. 
  \DP doesn't necessarily have to work at the level of files; one could certainly use prediction models at the level of modules, or even at the level of methods. To our
  knowledge, no one has done purely statistical models to predict defects at a line-level, and this constitutes a novel aspect of our work. 
  While earlier work evaluated models using IR measures such as precision, recall and F-score, more recently non-parametric methods such as AUC and AUCEC have
 gained in popularity.

 \subsection{Static Bug Finders} 
 The core idea of static bug finding is to develop an algorithm that automatically finds likely locations of known categories of defects in code. Some use heuristic pattern-matching; others are sophisticated algorithms that compute well-defined semantic properties over abstractions of programs carefully designed to accomplished specific speed-\emph{vs}-accuracy tradeoffs in detecting certain categories of bugs. The former tools include FindBugs and PMD, which we studied; the latter includes tools like ESC-Java~\cite{flanagan2002extended}. The former category can have both false positives and negatives. The over-riding imperative in the latter approach is to never falsely certify a program (that actually has \eg  memory leak bugs) to be bug-free; typically however, false positives can be expected. 
 
 The field has advanced rapidly, with many developments; researchers identify new categories of defects, and seek to invent clever methods to find these defects efficiently, either heuristically or though well-defined algorithms and abstractions. Since neither method is perfect, the actual effectiveness in practice is an empirical question. Since our goal here is just to compare \sta and \nbf, we  refer the reader for a more complete discussion of related work regarding \sta and their evaluation to Rahman \etal~\cite{rahman2014comparing}. 
  
 \subsection{Grammatical Error Correction in NLP}

Grammatical error correction is an important problem in natural language processing (NLP), which is to identify grammatical errors and provide possible corrections for them. The pioneering work on grammatical error correction was done by Knight and Chander~\cite{Knight:1994:AAAI} on article errors. Along the same direction, researchers haver proposed different classifiers with better features for correcting article and preposition errors~\cite{Han:2006:NLE,Tetreault:2008:COLING,Gamon:2010:NAACL,Dahlmeier:2011:ACL}. However, the classifier approaches mainly focus on identifying and correcting specific types of errors (e.g. preposition misuse). To approach this problem, some researchers have begun to apply the statistical machine translation approach to error correction. For example, Park and Levy~\cite{Park:2011:ACL} model various types of human errors using a noisy channel model, while Dahlmeier and Ng~\cite{Dahlmeier:2012:EMNLP} describe a discriminative decoder to allow the use of discriminative expert classifiers.
 
There is one fundamental difference between grammatical error correction in natural languages and defect localization in programming languages. Natural languages are close-vocabulary (i.e. have limited number of vocabulary), thus lead to limited types of grammatical errors (e.g. articles, prepositions, noun number) with enumerable corrections (e.g. possible article choices are {\em a/an}, {\em the}, and the empty article $\epsilon$). In contrast, programming languages are open-vocabulary (e.g. programmers could arbitrarily construct new identifiers). Therefore, the defects in programming languages are more flexible and thus harder to localize. Based on the observation that software corpora are highly repetitive~\cite{Hindle:2012:ICSE} and localized~\cite{Tu:2014:FSE}, we exploit a cache language model~\cite{Tu:2014:FSE} to locate the defects that are not {\em natural} in the sense that the sequences of code are not observed frequently either in the training code repository or in the local file.

\section{Conclusion}
\label{sec:conclusion}
The repetitive, predictable nature (``naturalness'') of code suggests that code that is improbable (``unnatural'') might be wrong. We investigate this intuition by 
using entropy, as measured by statistical language models, as a way of measuring unnaturalness. 

We find that unnatural code is more likely to be implicated in a bug-fix commit. We also find that buggy code tends to become more natural when repaired. We then turned to applying entropy scores to defect \emph{prediction} and find that, when adjusted for syntactic variances as well as syntactic variance in defect occurrence, it is about as cost-effective as the commonly used
static bug-finders PMD and FindBugs. 

Finally, applying the (deterministic) ordering of entropy scores to the warnings produced by these static bug-finders produces the most cost-effective method. These findings suggest that entropy scores
are a useful adjunct to defect prediction methods. The findings also suggest that certain kinds of automated 
search-based bug-repair methods might do well to have the search in some way influenced by language models.

\newpage
\newpage 

\balance
\bibliographystyle{abbrv}
\bibliography{main}  
%
%


\end{document}